\documentclass[
reprint,
showpacs,amsmath,amssymb,apl]
{revtex4-1}

\usepackage{textcase}
\usepackage{natbib}
\usepackage{graphics}
\usepackage{dcolumn}
\usepackage{bm}

\begin{document}

\preprint{APS/123-QED}

\title{High intrinsic energy resolution photon number resolving detectors}

\thanks{This research receives funding from the European Community’s Seventh Framework Programme, ERA-NET Plus, under Grant Agreement No. 912/2009/EC}%

\author{L. Lolli, E. Taralli, C. Portesi, E. Monticone}
\author{ M. Rajteri}%

\affiliation{INRIM - Instituto Nazionale di Ricerca Metrologica\\
 Strada delle Cacce 91, 10135 Torino, Italy (IT)}

\begin{abstract}
 
Transition Edge Sensors (TESs) are characterized by the intrinsic figure of merit to resolve both the energy and the statistical distribution of the incident photons. These properties lead TES devices to become the best single photon detector for quantum technology experiments. For a TES based on titanium and gold has been reached, at telecommunication wavelength, an unprecedented intrinsic energy resolution (0.113 eV). The uncertainties analysis of both energy resolution and photon state assignment has been discussed. The thermal properties of the superconductive device have been studied by fitting the bias curve to evaluate theoretical limit of the energy resolution.

\end{abstract}



\maketitle

Transition-Edge Sensors (TESs) are very versatile superconducting devices \cite{CPD}, that can be used to detect radiation in a wide energy range: from gamma-rays \cite{karasik} to visible \cite{cabrera} and sub-millimeter \cite{sub-milli}. In principle, they operate as thermometers: the absorption of incident photons heats the device, which is biased between the superconducting and the normal phase, causing a change in resistance; this variation is proportional to the photon energy or, for multi-photons at fixed wavelength, to the photon number. This intrinsic capability to measure the energy of the absorbed photons and the possibility to resolve the number of incident photons distinguishes single photon detectors from photon number resolving (PNR) devices.

The PNR property is useful in many scientific fields like astronomy, material analysis, quantum information, quantum computation and metrology. All these scientific areas need detectors with high energy resolution ($\Delta E$) and high quantum efficiency (QE). In refs.~\cite{irwin, kozo} a detailed analysis shows that the energy resolution strongly depends on temperature since it affects the heat capacity, the thermal conductance and the thermal noise. \\
The theoretical energy resolution of a generic calorimeter could be expressed as \cite{CPD}
~
\begin{eqnarray}
\Delta E_{rms} =\left( \sqrt{\int^\infty_0 \frac{4}{S_P(\omega)}~ \textrm{d}f} \right)^{-1}
\end{eqnarray}
~
where $\omega=2\pi f$ and $S_{P}(\omega)$ is the power spectral density of the total noise: for the ideal calorimeter it only includes the contributions of  phonon noise, detector Johnson noise and load resistor Johnson noise. In the approximation of dominant gaussian noise sources, neglecting the noise contributions from the other sources, and integrating over all the frequency band, the full width at half maximum theoretical energy resolution of a TES calorimeter can be derivate by the ``Figueroa model'' \cite{miller}
~
\begin{widetext}
\begin{equation}\label{DeltaE_teo}
\Delta E_{the} = 2\sqrt{2\ln(2)}  \sqrt{4 k_b T_0^2\frac{C_e}{\alpha} \sqrt{\frac{n}{\phi} (1+r) \left\{ \frac{1}{2} \left( 1+\frac{T_s^2}{T_0^2}\right)
+ \frac{n}{\alpha^2 \phi}  \left[1+r \left(1+\frac{\alpha \phi}{n}\right)^2\right]\right\} }} 
\end{equation}
\end{widetext}
~
where $\phi=1- (T_s/T_0)^n$ and $r=R_s T_s / (R_0 T_0)$.  $R_0$ is quiescent operating resistance of the TES and $R_s$ is the shunt resistance. The thermal behavior of the calorimeter is dominated by the decoupling between the electron and phonon system and described by the exponent factor $n$ of the thermal conductance equation $G(T)$. In this model, the main heat capacity is issued from the electronic system $C_e$ at temperature $T_0$. The thermal conductance $G$ links the electronic system to the phonon one, which is assumed at the substrate temperature $T_s$.  The detector thermal sensitivity is $\alpha = (T/R) (\textrm{d}R/\textrm{d}T)$, which results bigger than $n$, in the strong ETF limit, while $T_0>T_s$.

The electronic heat capacity plays a crucial role to determine the device behavior. A heat capacity that is too large will lead a temperature change during an event that will be too small to be accurately measured above the device noise fluctuations. A heat capacity that is too small will cause the temperature excursion during an event to exceed the width of the superconducting transition, causing device saturation. 

At INRIM, the best compromise has been obtained producing TESs based on titanium and gold films, a structure commonly used in metalization of electronic devices and easy to reproduce by traditional deposition techniques \cite{chiara}. The PNR capability of Ti/Au TESs has already been studied and described in \cite{LTD}, where up to twenty-nine photons have been resolved without reaching the detector saturation. Even the possibility of having fast recovery time lower than 200~ns has been demonstrated with this kind of TESs \cite{asc12, tesi}. A quantum characterization of superconducting photon counters \cite{povm} and a self consistent, absolute calibration technique for photon number resolving detectors \cite{cal} have been recently demonstrated. 

In this letter we discuss on an unprecedented energy resolution obtained in the telecom spectral range, paying attention on the main contributions of uncertainty, for both energy resolution and photon state assignment. The theoretical limit of the energy resolution is also evaluated knowing the necessary parameters by fitting the detector bias curve. 

The Ti/Au detector developed for the above aim was based on a 45~nm thickness layer of titanium deposited over a 45~nm gold film. The substrate was a silicon wafer covered by silicon nitride layer, the device active area was a square of 10~$\mu$m side. By exploiting the 4-wires technique and a resistance bridge, one of this TES has been measured, finding the normal resistance $R_n\simeq0.45~\Omega$ and the critical temperature $T_c\simeq106$~mK. 

\begin{figure}
\includegraphics{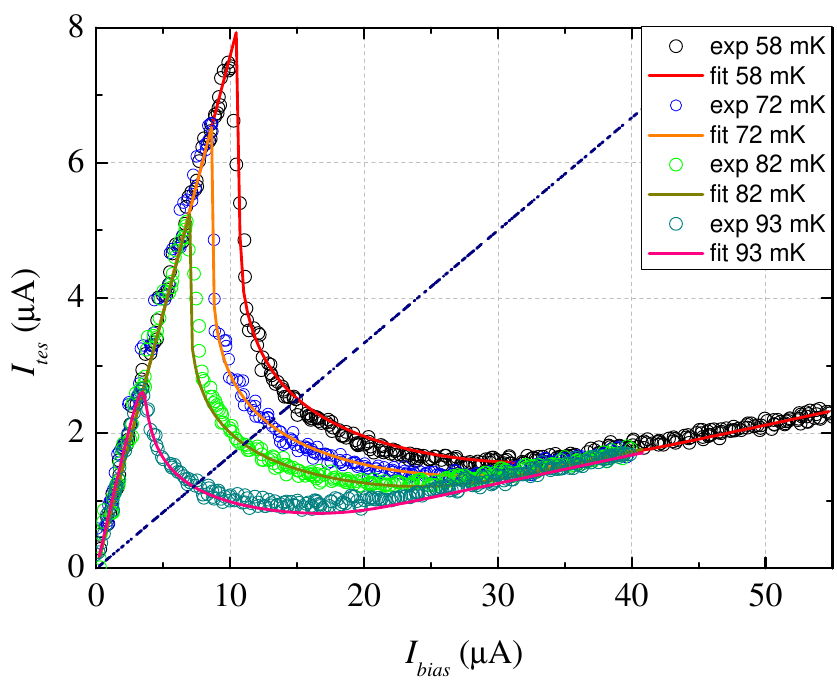}
\caption{\label{due} TES bias curves $I_{tes}$ vs.~$I_{bias}$ at different bath temperatures (58~mK, 72~mK, 82~mK and 93~mK): circles are the experimental data and solid lines are their respective fits. Dashed line indicates the resistance (0.1~$\Omega$) corresponding to the working point into the transition region.}
\end{figure}

To estimate the device properties, the detector was voltage biased with a shunt resistor $R_s = 22$~m$\Omega$ and readout with a dc-SQUID array current amplifier, with an input coil of $4$~nH. At several bath temperatures, TES bias curves ($I_{tes}$ vs.~$I_{bias}$) have been acquired: Fig.~\ref{due} shows 4 of these curves (circle) with their corresponding fits (solid lines). 
~

The fit function is obtained by modeling the TES resistance by an hyperbolic tangent function, as expression of current and temperature \cite{bay}, 
~
\begin{eqnarray}\label{R}
R(T,I)= \frac{R_N}{2} \left[ 1+ \tanh\left( \frac{T-T_c + \xi \cdot I}{D}\right) \right]
\end{eqnarray}
\\
and imposing the equilibrium state at the system of differential equations:
~
\begin{eqnarray}\label{RR}
\left\{
\begin{array}{l}
C_e\dot T=IR^2(T,I) - k (T^n-T_s^n) \\
L\dot I = I_{bias} R_s - I [R_p + R_s + R(T,I)] 
\end{array}
\right.
\end{eqnarray}
\\
The free parameters are $\xi$, which considers the current dependence of the resistance, and $k$ related to the thermal conductance $G$. $D$ value is chosen to accurately model the transition width at negligible current.

To operate the TES as a photon number detector, the bath temperature was adjusted at $T_s = 58$~mK and the working point was chosen setting $I_{bias} = 15~\mu$A, which corresponds to bias the device at 22~$\%$ of its normal resistance. The fit of this bias curves converges to $D=2.8$~mK, $\xi=1570$~K/A and $k=703$~nW/K$^5$, with $n=5$  and a parasitic resistance $R_p= 7$~m$\Omega$.

From these values it is possible now to calculate the thermal conductance $G=nkT_c^{n-1}\simeq44$~pW/K 
and the thermal sensitivity at the working point $\alpha= 23$. These numbers are consistent with those estimated by our previous impedance measurements on similar Ti/Au TESs \cite{ET, EPJ} ($10\leq\alpha\leq50$ with $n=5$), confirming the goodness of the two methods to study the TES characteristics. 

From literature data \cite{data}, the TES heat capacity can be estimated in $0.35$~fJ/K. With the obtained $\alpha$, the theoretical energy resolution (Eq.~(\ref{DeltaE_teo})) becomes $\Delta E_{the}\simeq0.064$~eV. 

\begin{figure}
\includegraphics{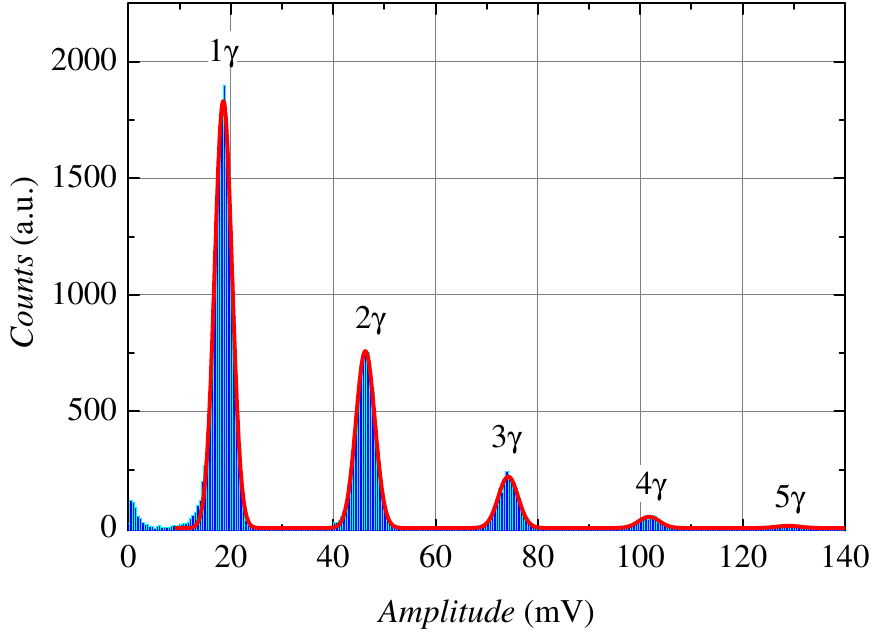}
\caption{\label{uno} Amplitude histogram of Wiener filtered signals (blue bars) shows from one ($1\gamma$) up to five ($4\gamma$) photon states. the FWHM energy resolution of the data fit (red line) is $\Delta E=~(0.113\pm0.001)~\textrm{eV}$.}
\end{figure}

The TES was irradiated as in the experimental setup described in \cite{tesi, V-g}: the distance between the tip of the spherical lensed fiber and the TES was fixed to obtain the minimum beam waist on the detector surface. 
Fig.~\ref{uno} shows the pulse amplitude histogram acquired with photons of energy $E_\gamma= 0.79$~eV, obtained by an attenuated pulsed coherent source at 1570~nm. The histogram of Fig.~\ref{uno} has four peaks: from left to right, from the single photon state ($1\gamma$) up to four ($4\gamma$) detected photon state. The peak of $0\gamma$ has been isolated and neglected using the cross-correlation information during the reduction noise process. The amplitude signals were analyzed via Wiener filtering process \cite{Diego} to increase the signal-to-noise ratio: it could improve of a factor 2 the energy resolution. The histogram data were then fitted considering free Gaussian distributions for each photon state (continuous line of in Fig.~\ref{uno}). The full width at half maximum (FWHM) energy resolution is related to the single photon peak standard deviation $\sigma_1$ by the formula \cite{tesi}:
~
\begin{eqnarray}\label{DeltaE_exp}
\Delta E = 2\sqrt{2~ln(2)} \cdot \frac{\sigma_1~E_\gamma}{x_{2\gamma} - x_{1\gamma}}
\end{eqnarray}
~
where denominator is the difference between the amplitude centers of the first two photon states. From the histogram fit, FWHM energy resolution was estimated in $\Delta E=(0.113\pm0.001)$~eV. 

The difference with the theoretical value could be due to a not perfect alignment of the optical fiber over the TES. This means photons are absorbed by the aluminum wires, deposited on the edge of the detector active area, or on the substrate near the TES, reducing the signal-to-noise ratio and consequently the intrinsic energy resolution of the detector. Nevertheless this energy resolution has never been reached in this spectral range. Furthermore, to compare the experimental energy resolution with the theoretical value, one should also considers the collection efficiency, a correction factor due to the real energy of the absorbed photon and the corresponding experimental energy computed from the measured pulse \cite{miller}. At telecom wavelengths, Ti/Au TESs have already shown collection efficiency values ranging between $\varepsilon =0.90 \div 0.94$ \cite{LTD, V-g}, in this work the value is even improved reaching $\varepsilon=0.98$.

To obtain the $\Delta E$ variance, Eq.~(\ref{DeltaE_exp}) has been propagated at linear order, with the energy of the incident photon $E_\gamma$ expressed in term of light speed in vacuum $c$, Planck constant $h$, laser source wavelength $\lambda$ and converted in eV energy unit ($\epsilon$ factor) \cite{tesi, DiegoT}: 

\begin{widetext}
\begin{eqnarray}\label{sigma_delta_E}
~
\begin{array}{rl}
\sigma^2_{\Delta E} = & \left | \frac{\partial \Delta E}{\partial h} \right |^2 \sigma_h^2 + \left | \frac{\partial \Delta E}{\partial \lambda} \right |^2 \sigma_\lambda^2 + \left | \frac{\partial \Delta E}{\partial \epsilon} \right |^2 \sigma^2_{\epsilon} +  \left | \frac{\partial \Delta E}{\partial \sigma_1} \right |^2 \sigma^2_{\sigma_1} + \left | \frac{\partial \Delta E}{\partial x_{2\gamma}} \right |^2 \sigma^2_{x_{2\gamma}} + \left | \frac{\partial \Delta E}{\partial x_{1\gamma}} \right |^2 \sigma^2_{x_{1\gamma}} = \\
~
= & \left | \frac{E_\gamma ~ \sigma_1}{h ~ (x_{2\gamma}-x_{1\gamma})} \right |^2 \sigma^2_h + \left |- \frac{E_\lambda ~ \sigma_1}{\lambda ~ (x_{2\gamma}-x_{1\gamma})} \right |^2 \sigma^2_\lambda +\left |- \frac{E_\lambda ~\sigma_1}{\epsilon ~ (x_{2\gamma}-x_{1\gamma})} \right |^2 \sigma^2_\epsilon + \left | \frac{E_\lambda}{x_{2\gamma}-x_{1\gamma}} \right |^2 \sigma^2_{\sigma_1} + \\
~
+ & \left |- \frac{E_\lambda ~ \sigma_1}{(x_{2\gamma}-x_{1\gamma})^2} \right |^2 \sigma^2_{x_{2\gamma}} + \left | \frac{E_\lambda ~ \sigma_1}{(x_{2\gamma}-x_{1\gamma})^2} \right |^2 \sigma^2_{x_{1\gamma}} 
~
\end{array} 
~
\end{eqnarray}
\end{widetext}

Fundamental constant values and their uncertainty were taken from CODATA 2010 \cite{codata} and led negligible contributions to $\sigma^2_{\Delta E}$; photon wavelength $\lambda$ (and so $E_\gamma$) magnitude and its relative standard uncertainty $u_r(\lambda)=1.3\times10^{-3}$ are given by source dealer. For the signal amplitude relative uncertainties we have to consider the contributions due to both the fit analysis and the oscilloscope accuracy. The first one was $ u_r(x_{1,\ 2})\simeq1\times10^{-3}$ for both the first two peaks; the second one was weighted with the number of signals carrying 1 and 2 photons: e.g.~20000 for the first state, becoming negligible contributions. The main fit contribution, to the relative uncertainty budget, was due to the FWHM of single photon peak:  relative standard uncertainty $u_r(\sigma_1)= 9\times10^{-3}$. This analysis leads to an overall standard relative uncertainties of $\Delta E$ equal to $u_r(\Delta E)=9\times10^{-3}$. The $\Delta E$ uncertainty budget is summarized in Table~\ref{tab1}.

\begin{table}[!h]
\caption{\label{tab1}%
Main contributions to the uncertainty budget of the FWHM $\Delta E$ determination.
}
\begin{ruledtabular}
\begin{tabular}{lcr}
\textrm{Quantity}&
\textrm{~} &
\textrm{Relative Uncertainty}\\
\colrule

Photon Wavelength & $\lambda$  & $1.3\times10^{-3}$ \\
$1\gamma$ amplitude & $x_{1\gamma}$ &  $1.0\times10^{-3}$ \\
$1\gamma$ peak standard deviation & $\sigma_1$ & $9\times10^{-3}$ \\
$2\gamma$ amplitude & $x_{2\gamma}$ & $9\times10^{-4}$ \\

\end{tabular}
\end{ruledtabular}
\end{table}

To assign the correct number of photons to a pulse with an uncertainty lower than $1\%$, two adjacent Gaussian distributions of photon state, with same amplitude, should overlap at $3\sigma$ from their mean value \cite{ML}. It is commonly known as the source emission has some peculiar statistics, as it will be observable in the detection statistics, especially by using linear devices like TESs \cite{povm}. Hence for commercial attenuated diode laser sources the photon emissions have Poissonian distributions, as the amplitude of the photon states, like in Fig.~\ref{uno}, and the `$3\sigma$ distance' condition is sufficient for an assignment uncertainty lower than $1\%$. However, in the case of sources based on parametric down conversion or on vacancy centers in nano-diamonds, where the aim is the single photon state, the amplitude difference is so high that the `$3\sigma$ distance' condition could not be always sufficient (f.e.~see Fig.~5 in \cite{NV}).  For this reason, the assignment uncertainty of the correct number of photons to a pulse is peculiar not only of detector properties but also of source's.\\ Table~\ref{tab2} summarizes the mean values and the standard deviation values of the histogram of Fig.~\ref{uno}. In this distribution, fixing the thresholds at half between the mean values, the single photon state assignment has an outstanding uncertainty lower than $10^{-13}$. This means that with our detector the photon state assignment uncertainty is still below $1\%$ until the amplitude ratio between two adjacent Gaussian peaks reaches 5 part in $10^{11}$  (considering $1\gamma$ and $2\gamma$ states).

\begin{table}[!h]
\caption{\label{tab2}%
Fit data of Fig.~\ref{uno} for the uncertainty estimation (k=1) of the photon number.
}
\begin{ruledtabular}
\begin{tabular}{lcr}
\textrm{Fit Parameter} &
\textrm{(mV)} &
\textrm{$U$ (mV) }\\
\colrule

$\sigma_1$ &  $1.69$ & $0.015$ \\
$x_{1\gamma}$ & $18.54$ & $0.01$\\
$\sigma_2$ & $1.85$ & $0.02$\\
$x_{2\gamma}$ & $46.30$ & 0.02\\
$\sigma_3$ & $1.93$ & $0.08$ \\
$x_{3\gamma}$ & $74.19$ & $ 0.08$\\
$\sigma_4$ & $2.0$ & $0.4$ \\
$x_{4\gamma}$ & $101.7$ & $ 0.4$\\

\end{tabular}
\end{ruledtabular}
\end{table}

In this acquisition, the mean photon number per pulse was $\sim$1 at 9~kHz of repetition rate, resulting in a photon flux of about $9\times10^3$~photons/s (0.1 fW). Hence, the possibility of having a device able to work from low photon flux regime (1~photons/s) to flux measurable by conventional photon avalanche detector (up to $6\times10^4$~photons/s has been measured with this TES), with an uncertainty of photon state discrimination lower than $1\%$, is feasible.

In \cite{cal} we proposed an innovative technique to absolutely calibrate photon number resolving detectors, using a pulsed heralded photon source based on parametric down conversion. By exploiting this method, we evaluated the efficiency obtaining $30\%\pm5\%$. The consistent improve with respect to data in ref.~\cite{cal} regards the reduction of the geometrical and optical losses inside the refrigerator. Since the absence of any kind of structure able to reduce the natural reflection of the metal film, the efficiency upper limit of these Ti/Au TESs is $\sim 50\%$.

In summary we have described an approach to evaluate the properties of TES by exploiting their bias curve and demonstrated the possibility of reaching high values of energy resolution for Ti/Au TES in the telecom spectral range. For the first time, we present and discuss the measurement uncertainty of the intrinsic detector resolution and of assignment of the correct photon state number.


\thebibliography{apssamp}


\bibitem{CPD} K.D.~Irwin and G.C.~Hilton, in {\em Cryogenic Particle Detection}, Topics in Applied Physics, Vol.~99, edited by C.~Enss (Springer-Verlag Berlin Heidelberg 2005), Chap.~3, pp.~63--149.  

\bibitem{karasik} J.~Wei, D.~Olaya, B.S.~Karasik, S.V.~Pereverzev, A.V.~Sergeev and M.E.~Gershenson, {\em Nat.~Nanotechnol.}, {\bf 3}, 496 (2008).

\bibitem{cabrera} B.~Cabrera, R.M.~Clarke, P.~Colling, A.J.~Miller, S.~Nam, and R.W.~Romani, {\em Appl.~Phys.~Lett.}, {\bf 73}, 735 (1998).

\bibitem{sub-milli} R.~O'Brient, P.~Ade, K.~Arnold, J.~Edwards, G.~Engargiola, W.L.~Holzapfel, A.T.~Lee, M.J.~Myers, E.~Quealy, G.~Rebeiz, P.~Richards, and A.~Suzuki, {\em Appl.~Phys.~Lett.} {\bf 102}, 063506 (2013).

\bibitem{irwin} K.D.~Irwin, 
{\em Appl.~Phys.~Lett.}, {\bf 66}, 1998 (1995).

\bibitem{kozo} A.G.~Kozorezov, J.K.~Wigmore, D.~Martin, P.~Verhoeve, and A.~Peacock, 
{\em Appl.~Phys.~Lett.}, {\bf 89}, 223510 (2006).

\bibitem{miller} A.J.~Miller, 
Ph.D.~Thesis, Stanford University (2001).

\bibitem{chiara} C.~Portesi, E.~Taralli, R.~Rocci, M.~Rajteri, and M.~Monticone, 
{\em J.~Low.~Temp.~Phys.}, {\bf 151}, n.~1--2, 261 (2008).

\bibitem{LTD} L.~Lolli, E.~Taralli, and M.~Rajteri, 
{\em J.~Low.~Temp.~Phys.}, {\bf 167}, 803 (2012).

\bibitem{asc12} L.~Lolli, E.~Taralli, M.~Rajteri, T.~Numata, and D.~Fukuda, 
{\em IEEE Trans.~Appl.~Supercond.},  {\bf 23}, 2100904 (2013).

\bibitem{tesi} L.~Lolli, 
Ph.D.~Thesis, Politecnico di Torino (2012).

\bibitem{povm} G.~Brida, L.~Ciavarella, I.P.~Degiovanni, M.~Genovese, L.~Lolli, M.G.~Mingolla, F.~Piacentini, M.~Rajteri, E.~Taralli, and M.G.A.~Paris, 
{\em New.~J.~Phys.}, {\bf 14}, 085001 (2012).

\bibitem{cal} A.~Avella, G.~Brida, I.P.~Degiovanni, M.~Genovese, M.~Gramegna, L.~Lolli, E Monticone, C.~Portesi, M.~Rajteri, M.L.~Rastello, E.~Taralli, P.~Traina and M.~White, 
{\em Optics Express}, {\bf 19}, n.~23, 23249 (2011).

\bibitem{bay} T.~J.~Bay, 
Ph.D.~Thesis, Stanford University (2007).

\bibitem{ET} E.~Taralli, C.~Portesi, L.~Lolli, E.~Monticone, M.~Rajteri, I.~Novikov and J.~Beyer, 
{\em  Supercond.~Sci.~Technol.}, {\bf 23}, 105012 (2010).

\bibitem{EPJ} E.~Taralli, C.~Portesi, L.~Lolli E.~Monticone, M.~Rajteri, I.~Novikov and J.~Beyer, 
{\em Eur.~Phys.~J.~Plus.}, {\bf 127}, n.~2, 13 (2012).

\bibitem{data} Yong-Hamb Kim, Ph.D.~Thesis, Brown University (2004).

\bibitem{V-g}L.~Lolli, E.~Taralli, C.~Portesi, D.~Alberto, M.~Rajteri and E.~Monticone, 
{\em IEEE Trans.~Appl.~Supercond.}, {\bf 21}, n.~5, 215 (2011).

\bibitem{Diego} D.~Alberto, M.~Rajteri, E.~Taralli, L.~Lolli, C.~Portesi, E.~Monticone, Y.~Jia, R.~Garello and M.~Greco, 
{\em IEEE Trans.~Appl.~Supercond.}, {\bf 21}, n.~3, 285 (2011).

\bibitem{DiegoT} D.~Alberto, 
Ph.D.~Thesis, Politecnico di Torino and CERN, (2011-008).

\bibitem{ijqi} L.~Lolli, G.~Brida, I.P.~Degiovanni, M.~Gramegna, E.~Monticone, F.~Piacentini, C.~Portesi, M.~Rajteri, I.~Ruo Berchera, E.~Taralli and P.~Traina, 
{\em Int.~Journ.~Quant.~Inf.}, {\bf 9}, Suppl, 105012 (2011).

\bibitem{codata} P.J.~Mohr, B.N.~Taylor and D.B.~Newell, in {\em The 2010 CODATA Recommended Values}, CODATA Recommended Values of the Fundamental Physical Constants: 2010 (National Institute of Standards and Technology, Gaithersbug, Maryland, 2012), Chap.~XIV, pp.~64--81.

\bibitem{ML} J.C.~Zwinkels, E.~Ikonen, N.P.~Fox, G.~Ulm, and M.L.~Rastello, 
{\em Metrologia}, {\bf 47}, R15 (2010).

\bibitem{NV} W.~Schumnk, M.~Gramegna, G.~Brida,  I.P.~Degiovanni, M.~Genovese, H.~Hofer, S.~Kuck, L.~Lolli, M.G.A.~Paris, S.~Peters, M.~Rajteri, A.M.~Racu, A.~Ruschhaupt, E.~Taralli and P.~Traina, {\em Metrologia}, {\bf 49}, S156 (2012).
\end{document}